\documentclass[aps,eqsecnum,showpacs,twocolumn,floatfix,superscriptaddress]{revtex4}
\usepackage{graphicx}
\usepackage{amsmath}
\usepackage{amsfonts}
\usepackage{amssymb}
\usepackage{bm}
\newcommand{\be}{\begin{eqnarray}}
\newcommand{\ee}{\end{eqnarray}}
\newcommand{\ba}{\begin{array}}
\newcommand{\ea}{\end{array}}

\newcommand{\e}{\epsilon}
\newcommand{\ben}{\begin{eqnarray*}}
\newcommand{\een}{\end{eqnarray*}}
\def\beq{\begin{equation}}
\def\eeq{\end{equation}}
\begin{document}

\title{Classical and quantum anomalous diffusion in a system of
2$\delta$-kicked Quantum Rotors}
\author{Jiao Wang}
\affiliation{Temasek Laboratories,National University of
Singapore,117542 Singapore.}
\affiliation{Beijing-Hong
Kong-Singapore Joint Center for Nonlinear and Complex Systems
(Singapore), National University of Singapore,117542 Singapore.}
\author{Antonio M. Garc\'{\i}a-Garc\'{\i}a}
\affiliation{Physics Department, Princeton University, Princeton,
New Jersey 08544, USA}
\affiliation{The Abdus Salam International Centre for Theoretical
Physics, P.O.B. 586, 34100 Trieste, Italy}

\date{\today}

\begin{abstract}
We study the dynamics of cold atoms subjected to {\em pairs} of
closely time-spaced $\delta$-kicks from standing waves of light. The classical phase space of this system is partitioned into
momentum cells separated by trapping regions. In a certain range of parameters it is shown
that the classical motion is
well described by a process of anomalous diffusion. We investigate
in detail the impact of the underlying classical anomalous
diffusion on the quantum dynamics with special emphasis on the
phenomenon of dynamical localization.  Based on the study of the
quantum density of probability, its second moment and the return
probability we identify a region of weak dynamical localization
where the quantum diffusion is still anomalous but the diffusion
rate is slower than in the classical case.  Moreover we
examine how other relevant time scales such as the
quantum-classical breaking time or the one related to the beginning of full dynamical
localization are modified by the classical anomalous diffusion.
Finally we discuss the relevance of our results for the
understanding of the role of classical cantori in quantum
mechanics.
\end{abstract}

\pacs{32.80.Pj, 05.45.Mt, 05.60.-k}

\maketitle

\section{Introduction}
The effect of the underlying classical dynamics on the quantum
motion has been a recurrent topic of research since the early days
of the quantum theory. In recent years the use of experimental
techniques based on ultracold atoms in optical lattices
\cite{Raizen} has permitted to study in great detail the role of
classical mechanics in simple quantum systems. In these
experiments a very dilute almost free gas of atoms (Cs and Rb) is
cooled down to temperatures of the order of tens $\mu K$ and then
interacts with an optical lattice. In its simplest form, the
optical lattice consists of two laser beams prepared in such a way
that the resulting interference pattern is a stationary plane wave
in space. The laser frequency is tuned close to a resonance of the
atomic system in order to enhance the atom-laser coupling but not
too close to avoid spontaneous emission. In this limit the
laser-atom system can be considered as a point particle in a sine
potential, namely, the quantum pendulum. If the
laser is turned on and off in a series of short periodic pulses,
the resulting system is very well approximated by the so called
quantum kicked rotor (QKR) (see Ref. \cite{reviz} for a review)
extensively studied in the context of quantum chaos, \be {\cal H}=
\frac{p^2}{2} - K\cos(q)\sum_{n}\delta(t-Tn).\label{kr} \ee For
short time scales, quantum and classical motion agrees. However
quantum diffusion is eventually suppressed due to destructive
interference that localize eigenstates in momentum space. This
counter-intuitive feature, usually referred to as dynamical
localization \cite{dyn}, was fully understood \cite{fishman} after
mapping the kicked rotor problem onto a short range one
dimensional disordered system where localization is well
established.  The theoretical predictions of Ref. \cite{fishman}
were eventually confirmed experimentally \cite{Raizen} (see also
Ref. \cite{otherexp}) by using the cold atoms techniques mentioned
previously. The standard kicked rotor is thus an ideal 
candidate for the study of the quantum properties of one dimensional
systems whose classical motion is diffusive. A natural question to
ask is whether this analysis can be extended to other types of
(anomalous) diffusive motion.

For values of the kick strength $K$ in Eq.(\ref{kr}) sufficiently
small, the classical phase space is composed of chaotic and
integrable parts and, for certain initial conditions, the
classical motion is well described by a process of anomalous
diffusion. The quantum transport properties in systems with a
mixed phase space \cite{Christensen} 
depend strongly on the details of the Hamiltonian \cite{boh}.
Even for a given configuration many types of anomalous diffusion
are observed depending on the initial conditions or the time
scales studied \cite{Geisel}. This lack of universality makes it
difficult to precisely assess the effect of the underlying
classical anomalous diffusion on the quantum dynamics.

The situation is different if the smooth sinusoidal optical
potential is replaced by a potential with a logarithmic or
power-law singularity \cite{ant9}. The classical phase space is
homogeneous but still the classical motion is superdiffusive. In
the quantum realm, as a consequence of interference effects, the
particle still diffuses but at a slower rate. In fact, for certain
types of singularities, full dynamical localization  never occurs
and diffusion persists at all times. In other cases exponential
localization is eventually observed but  anomalous diffusion
different from the classical one is still observed for shorter
times. The classical density of probability $P(p,t)$ in this
region is accurately described by the solution of an anomalous
Fokker-Planck equation \cite{klafter}. We note that, unlike the
case of normal diffusion, the information obtained from the
knowledge of a few moments of the density of probability may not
be sufficient to fully characterize the classical motion
\cite{klafter}. Thus for a correct understanding of these systems
it is essential a detailed investigation of  $P(p,t)$.

The models studied in Ref. \cite{ant9} are non-KAM but classical
anomalous diffusion can also exist in KAM systems. One example is
the kicked rotor with a smooth potential but subjected to {\em
pairs} of closely time-spaced kicks: the $2\delta$-kicked rotor
($2\delta$-KR) \cite{Jones,Tania}.

The dynamics of this model in a certain region of parameters has
already been investigated in the literature: theoretically in Ref.
\cite{Tania} and experimentally in  Ref. \cite{Jones}. Unlike the
single kicked rotor the classical dynamics is strongly correlated.
The momentum space is divided into regions of fast momentum
diffusion separated by porous boundaries, ie narrow trapping
regions where classical trajectories `stick' for relatively long
periods.
The $2\delta$-KR trajectory spends considerable time trapped in a
cell before escaping to the next.

In this paper we provide a detailed account of the type of
classical and quantum anomalous diffusion associated with this
motion. We shall restrict ourselves to the one cell region,
namely, to time/momentum scales such that a particle initially
trapped  manages to escape and eventually reaches a new trapping
region. We aim a better understanding of how generic features of
the classical anomalous diffusion affect the quantum motion in a
KAM system. We restrict ourselves to a region of parameters such
that the dynamics is generic, namely, the classical phase space is
fully chaotic with no island of stability. However the
trapping-leaking mechanism of our model still causes strong
deviations from the single kicked rotor results. Other important
motivation to study the motion of a fully chaotic $2\delta$-KR is
that it mimics the effect of cantori in chaotic systems. Thus a
particle typically gets trapped in a cantorus for a long time
until it escapes to the chaotic sea. We argue the findings of this
paper shed light about the role of classical cantori in quantum
mechanics \cite{Maitra,weiss}.

The organization of the paper is as follows: in the next section
we introduce the model, review some of its more relevant dynamical
features and discuss the region of parameters to be studied. In
section III, we study the classical and quantum transport
properties. Among others we analyze the classical and quantum
density of probability, the classical-quantum breaking time and
the return probability. We aim to describe how dynamical
localization  arises in systems whose classical diffusion is
anomalous and how other relevant scales of the problem  such as
the quantum-classical breaking time are affected by the underlying
classical dynamics.

In summary, our main new results are:  1) In the region of
interest both classical and quantum dynamics is well characterized
by a process of anomalous diffusion. 2) We identify two routes to
dynamical localization as a function of $\hbar$: for $\hbar >
\hbar_c$ ($\hbar _c$ is function of the kick strength and the
separation between pairs of kicks) standard dynamical localization
occurs in the trapping region and the particle does not escape;
for $\hbar < \hbar_c$ the central part probability density is
exponential and almost time independent. However diffusion does
not stop since eventually the particle escapes from the trapping
region. True dynamical localization in this case occurs for time
scales much longer than the typical time to escape from the
trapping region. For intermediate times we find the quantum
diffusion is anomalous but slower than the classical one. 3) The
classical anomalous diffusion induces a fractional scaling with
$\hbar$ in different quantities of interest such as the
quantum-classical breaking time.  4) The $2\delta$-KR  can be used
as a simplified model to study the effect of cantori in classical
and quantum mechanics.

\section{The model}

We consider a system with a Hamiltonian corresponding to a
sequence of closely spaced pairs of kicks:\be {\cal H}=
\frac{p^2}{2} - K \cos x \sum_n \left[ \delta(t-nT)+
\delta(t-nT+\epsilon) \right]\nonumber, \ee where $\epsilon \ll T$
is a short time interval and $K$ is the kick-strength.

\subsection{Classical dynamics}
The classical map for the $2\delta$-KP is a straightforward
extension of the Standard Map:
\begin{eqnarray}
p_{n+1}=p_n + K\sin x_n; &\ & p_{n+2}=p_{n+1} + K\sin x_{n+1}
\nonumber \\
x_{n+1}=x_n + \epsilon p_{n+1}; &\ & x_{n+2}=x_{n+1} + \tau
p_{n+2}
\nonumber \\
\label{eq2}
\end{eqnarray}
where $\epsilon$ is a very short time interval between two kicks
in a pair and $\tau = T -\epsilon$ is the  (much longer) time interval
between the pairs.
Clearly, the limit $\epsilon=\tau$ or $0$ corresponds to the
Standard Map, which describes the classical dynamics of the
quantum kicked rotor:
\begin{eqnarray}
p_{i+1} &=& p_{i} + K \sin x_{i}, \nonumber \\
x_{i+1} &=& x_i + p_{i+1}. \label{eq3}
\end{eqnarray}
Particles with momenta $p_0 \simeq (2m+1)\pi/\epsilon$ and
$m=0,\pm 1, \pm 2,\cdots$ (relative to the optical lattice) are
confined in momentum trapping regions and absorb little energy;
conversely, particles prepared near $p_0 \simeq 2m\pi/\epsilon$
experience rapid energy growth up to localization. The basic
mechanism of trapping is fairly intuitive: atoms for which $p_0 =
(2m+1)\pi/\epsilon$ experience an impulse $K\sin x$ followed by
another one $\simeq K\sin(x + \pi)$ which in effect cancels the
first. Over time, however, there is a gradual de-phasing of this
classical `anti-resonant' process. A theoretical study
\cite{Jones} of the classical diffusion in this system found
anomalous momentum diffusion for any $p_0$ and intermediate time
scales. It was also observed long-ranged corrections to the
uncorrelated diffusion rate not present in the standard kicked
rotor. Inside the trapping region diffusion is normal but the
coefficient of diffusion $D$ has a dependence on $K$ as $D \sim
K^3$, similar to the one found in single kicked rotors in a region
of the classical phase space densely populated by cantori
\cite{weiss}. By contrast, if the phase space is fully chaotic it
is expected $D \sim K^2$. This suggests that our model may
reproduce to a good approximation the effect of cantori in a KAM
system.

The size in momentum space of the trapping region $\delta p$
strongly depends on the parameters $\epsilon,K$ defining the
model.

For an accurate analysis of the trapped region is necessary that:
i) the particle is initially trapped, ii) the particle dwells on
average a time long enough in this region before it escapes from
it. In Ref. \cite{Jones} it was concluded that this implies the
criterion $K \epsilon \ll 1$. However, if $K \epsilon$ is too
small, the phase space would be too regular.  Since we are
interested in generic feature of the motion, we take iii) $\tau
\gg 1$. This implies less correlations between successive space
positions of the particle. We thus restrict ourselves through the
paper to the range  $K \epsilon \ll 1$ and $\tau \gg 1$ with
initial conditions inside the trapping region.

\subsection{Quantum dynamics}

The time evolution operator for this system can be written as
\begin{eqnarray}
\hat{U}^{\epsilon} = e^{-i \frac{\tau \hat{p}^2}{2\hbar}}
e^{i\frac{K}{\hbar} \cos x} e^{-i \frac{\epsilon
\hat{p}^2}{2\hbar}} \ \e^{i\frac{K}{\hbar} \cos x}. \ \label{eq31}
\end{eqnarray}
In a basis of plane waves, $\hat{U}^{\epsilon}$ has matrix
elements
\begin{eqnarray}
U_{lm}^{\epsilon}=U_l^{free}. \ U_{lm}^{2-kick} = e^{-i
\frac{l^2\hbar\tau}{2}} \ i^{l-m} \nonumber  \\  \sum_k
J_{l-k}\left(K_{\hbar}\right) \ J_{k-m}\left(K_{\hbar}\right) \
e^{-i \frac{k^2\hbar \epsilon}{2}} \label{eq32}
\end{eqnarray}
where $K_{\hbar}= K/\hbar$ and $J_n(K_{\hbar})$ is integer Bessel
functions of the first kind. It is easy to see that
$U_{lm}^{2-kick}$ is invariant if the products $K_\epsilon=
K\epsilon$ and $\hbar_\epsilon= \hbar\epsilon$ are kept constant;
while the free propagator $U_l^{free}=e^{-i
\frac{l^2\hbar\tau}{2}}$ simply contributes a near-random phase.
Thus the results are quite insensitive to the magnitude of $\tau =
T-\epsilon$ provided that $\tau \gg 1$. We will stick to $K
\epsilon \ll 1$ and $\tau \gg 1$ in all numerical calculations.

The result in Eq. (\ref{eq32}) may be compared with the one-kick
map in Eq. (\ref{eq2})
\begin{eqnarray}
 U^{(0)}_{lm}= e^ {-i \frac{l^2 T \hbar}{2}} \
J_{l-m}\left(K_{\hbar}\right) . \label{eq33}
\end{eqnarray}
The one-kick matrix for the QKR has a well-studied band-structure:
since $J_{l-m}(x)\simeq 0$ for $|l-m| \gg x$, we can define a
bandwidth for $U^{(0)}$; i.e. $b= K_{\hbar}$ (this is strictly a
{\em half}-bandwidth) which is independent of the angular momenta
$l$ and $m$. However, this is {\em not} the case for the matrix of
$U^{\epsilon}$.

Assuming $|l-m|$ is small it was shown in Ref. \cite{Tania} that
\begin{equation}
U_{lm}^{\epsilon} \approx  e^{-i \Phi}  \  J_{l-m}\left[
2K_{\hbar} \cos \left ({l\hbar_\epsilon/2}\right) \right]
\label{eq34}
\end{equation}
where the phase $\Phi=(l^2T+ \epsilon l m + \epsilon l^2)\hbar/2+
\pi (l-m)/2$. Hence we infer a momentum dependent bandwidth,
$b(p)= 2K_{\hbar}\cos{(p \epsilon/2)}$.

While $U^{(0)}$ has a constant bandwidth, the bandwidth for the
matrix of $U^{\epsilon}$  oscillates with $l$ from a maximum value
$b_{max}=2K_{\hbar}$, equivalent to twice the bandwidth of
$U^{(0)}$, down to a minimum value $b_{min} \sim 0$.
 $U^{\epsilon}$ is thus partitioned
into sub-matrices of dimension $N ={2\pi}/{\hbar_\epsilon}$
corresponding precisely to the momentum cells of width $\Delta p =
N \hbar$ observed in experiments \cite{Jones}.

In this paper we aim to study both the evolution of an initially
given wave-packet and its low order moments. The structure of the
evolution operator $U^{\epsilon}$ allows an efficient and accurate
numerical calculation of these quantities. The action of the
operator $U^{\epsilon}$ on a quantum state can be decomposed into
four steps: two associated with the kicks, which are diagonal in
the space presentation; and two associated with the free rotations
between neighboring kicks, which are diagonal in the momentum
presentation. We thus take the space and momentum representations
alternatively to facilitate the calculations. The transformation
between the representations is efficiently carried out with the
fast Fourier transformation (FFT) algorithm. In our numerical
simulations we can set the size of the basis up to $2^{20}$ which
is big enough to guarantee a double precision accuracy.

\section{Results: Anomalous diffusion and dynamical localization}

In this section we investigate the classical and quantum dynamics
of the $2\delta$-KR. Our main motivation is to describe the effect
on the quantum motion of generic dynamical features of the
classical dynamics such as the trapping-escaping mechanism.  We
mainly focus on time/momentum scales such that the motion is
confined inside one cell.  Initial conditions are always chosen
within the trapping region. For a study in the multi-cell region
we refer to Ref. \cite{Tania,Jones}.

As was mentioned previously, we restrict ourselves to the window
of parameters $K \epsilon \ll 1$. In addition it is also imposed
$\tau \gg 1$ in order to remove correlations that make the
dynamics non generic. In the numerical calculations $\tau, K$, and
$\epsilon$ are fixed within the above limits.  Our main
conclusions do not depend on the specific value of the parameters.
Quantum effects are
investigated by varying $\hbar$.  
\begin{figure}
\includegraphics[width=.95\columnwidth,clip]{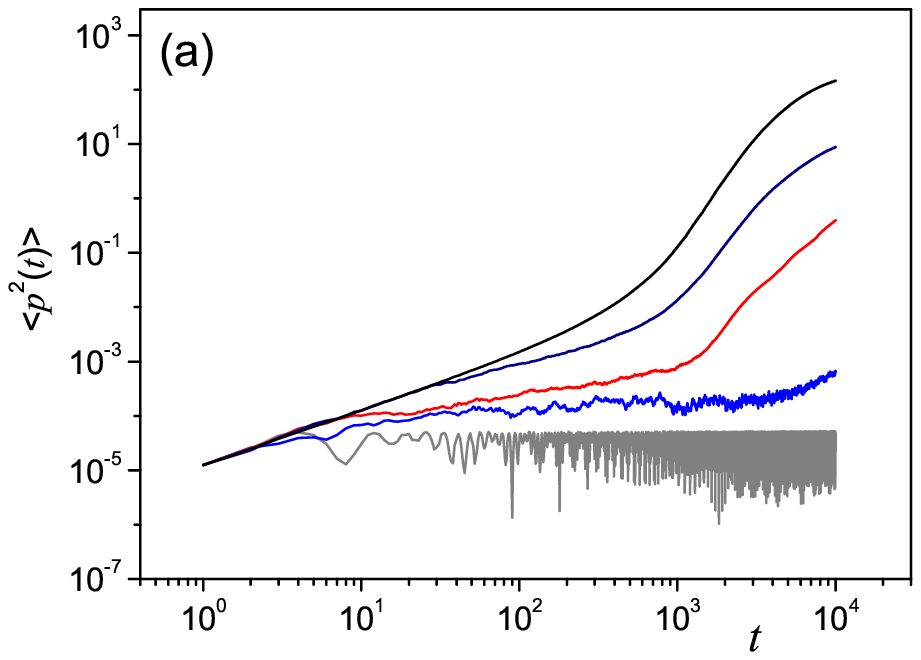}
\vspace{-.5cm}\label{fig1a}
\includegraphics[width=.95\columnwidth,clip]{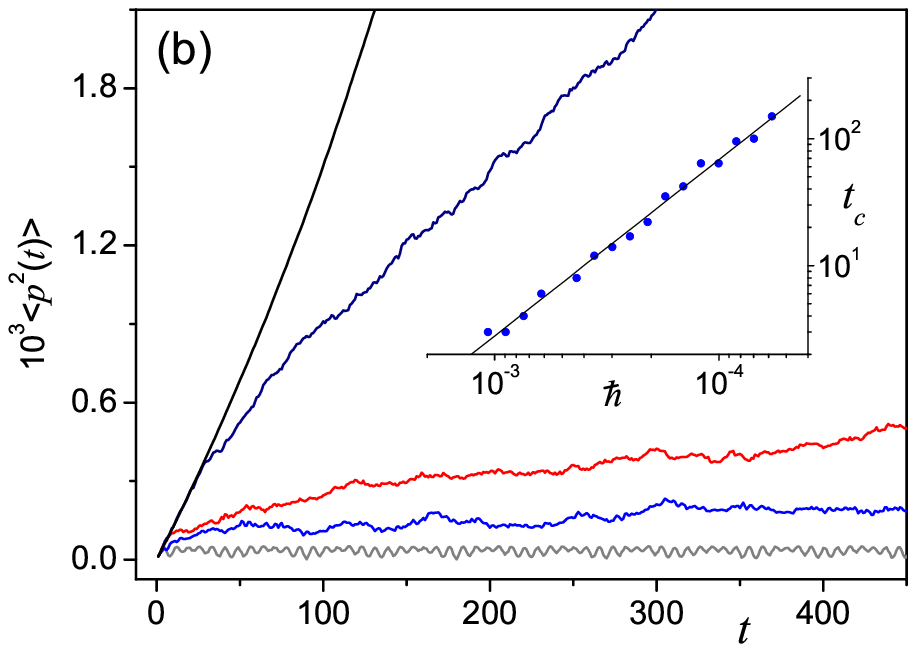}
\vspace{-.5cm}\label{fig1b} \caption{(Color online) Comparison of
quantum and classical energy diffusion (from bottom to top:
$\hbar=4\times 10^{-3},1\times 10^{-3},5\times 10^{-4}, 2.5\times
10^{-4}$ and classical respectively) for $K=0.2,\epsilon=0.25$ and
$\tau=10^4$. The two figures only differ in the scale of
time/momentum represented.  Fig.(a) shows the effect of leaking
from the trapping region in longer time scales while Fig.(b)
describes diffusion inside the trapping region. The inset of Fig.(b)
shows the dependence of quantum-classical breaking time $t_c$ on $\hbar$; the best fitting (thin
solid line) suggests that $t_c\sim \hbar ^ {-1.39\pm 0.05}$ for
$\hbar<10^{-3}$.  This is in contrast with the single kicked rotor
where $t_c \sim \hbar^{-2}$} \label{fig1}
\end{figure}

\subsection{Energy diffusion}
The classical dynamics in the trapping region was investigated
analytically in Ref. \cite{Jones}. It was found that energy
diffusion $\langle p^2(t) \rangle \sim D t$ increases linearly
with time. However, the dependence of $D \sim K^3$ on the kicking
strength is different from the single kicked rotor prediction $D
\sim K^2$ in the chaotic region but similar to the one observed if
the classical phase space is populated by cantori \cite{Fish}. This
suggests that our model captures generic features of the effect of
cantori in classical mechanics.

Eventually the particle reaches the edge of the trapping region
and leaks outside. Diffusion becomes then anomalous until the
particle gets trapped again. As is shown in Fig. 1, the parameters
$K, \epsilon$ and $\tau$ were chosen such that the typical time
for considerable leaking is $t_{leak} \approx 200 \gg 1 $. Before
this time the leakage is still possible but with a negligible
probability. As the particle approaches the outer boundaries of
the two cells that sandwich the trapping region (at $t\approx
6\times 10^3$), the diffusion begin to slow down. The
time unit we take in all figures is the number of the kick pairs.

In the quantum case the dynamics is richer. If we restrict
ourselves to the one cell region three time scales are
distinguished. In a first stage, $t < t_c$, the quantum averaged
energy agrees with its classical counterpart.  As a result of the
peculiarities of the classical dynamics, scaling with $\hbar$ of
the classical-quantum breaking time $t_c$ may be different from
$t_c \sim \hbar^{-2}$, the result for a single kicked rotor.
For longer times, $t_c < t < t_d$, interference effects start to
be important. As a consequence we expect the particle still
diffuses, $\langle p^2(t) \rangle \approx D_{quan} t^{\gamma}$,
but at a rate which decreases as $\hbar$ increases and it
is lower than the classical one; namely $\gamma \leq 1$ and
$D_{quan} \leq D_{clas}$. This is a novel regime caused by the
interplay of destructive interference and classical anomalous
diffusion.  This stage lasts up to a time $t_d$ in which full
dynamical localization occurs and diffusion is totally arrested.

Our numerical results confirm the above qualitative picture.
Results for quantum and classical energy diffusion $\langle p^2(t)
\rangle$ ($p=\hbar k$ for the quantum case) as a function of time
$t$ for $\tau = 10^4$ and $K \epsilon=0.05 \ll 1$ for different
$\hbar$'s are shown in Fig. 1. In the classical case $2\times
10^7$ initial conditions uniformly distributed along
$p_0=\pi/\epsilon$ (located at the center of the trapping region)
have been utilized for the ensemble average. The quantum initial
condition is set as the momentum eigenstate $|l_0\rangle$ with
$l_0$ the integer closest to $p_0/\hbar$.

We restrict ourselves to time/momentum scales such that the motion
is confined to one cell. We observe that for short times both
classical and quantum results coincide up to a certain breaking
time $t_c$ which has a fractional scaling with $\hbar$, different
from the single kick rotor case (see insert of Fig. 1b, where
$t_c$ is evaluated as the maximum time such that the deviation
between classical and quantum $\langle p^2(t)\rangle$ is below
$20\%$). For later times the quantum particle still diffuses
normally but with a smaller diffusion coefficient $D_{quan}$ (see
Tab. I). This is a direct consequence of destructive interference
effects. We relate this new region of weak dynamical localization
to the effect of classical trapping on the quantum dynamics. This
interpretation is reinforced by the observed dependence of
$D_{quan}$ on $\hbar$ in the trapping region (see Tab. I). For
$\hbar$ sufficiently small ($\hbar<10^{-4}$), $D_{quan}$ is close
to the classical prediction. For larger $\hbar$ we observe a
growing dependence on $\hbar$ though the exponent $\gamma \sim 1$
is not modified.  It decreases as $\hbar$ increases. This
situation lasts until a maximum $\hbar$, denoted as $\hbar_{max}$,
such that the breaking-time $t_c$ coincides with the time $t_d$
in which full dynamical localization starts. For the
parameters utilized $\hbar_{max} \approx 6\times 10^{-3}$.

We have observed an additional $\hbar$ scale, denoted as
$\hbar_c$, relevant for a precise characterization of the
trapping-leaking mechanism.  It is defined by the smallest $\hbar$
such that $t_d < t_{leak}$. For the parameters used
$\hbar_c\approx 10^{-3}<\hbar_{max}$. For $\hbar_c < \hbar <
\hbar_{max}$, $t_d < t_{leak}$, $D_{quan}$ is close to zero, and
the wave-packet is well localized in the trapping region. In this
range as $\hbar$ approaches $\hbar_{max}$, $t_d$ tends to $t_c$.
For $\hbar < \hbar_c$, the time scale $t_d$ related to the
dynamical localization increases dramatically. However quantum
energy diffusion is still linear in time (like the classical one)
up to $t_{leak}$. After $t\approx t_{leak}$ the classical particle
starts to leave the trapping region and the classical motion
become superdiffusive. The quantum dynamics for $\hbar< \hbar_{c}$
is also superdiffusive, but with a smaller leaking rate. The
smaller the $\hbar$, the closer the leaking rate to the classical
one. Eventually ($t\approx 6\times 10^3$) a new trapping region is
approached. As a consequence, both classical and quantum motions
are slowed down again.

We note that the region of weak dynamical localization
characterized by a linear time dependence of the energy evolution
is not present in the standard kicked rotor where there is no
intermediate region between classical diffusion and full quantum
dynamical localization. This is an indication that even though the
quantum diffusion is still normal inside the trapping region the
$2\delta$-KR is essentially different from the single kicked
rotor.

\begin{table}
\vspace{0.2cm}
\begin{tabular}{cccc}
  \hline
  $\hbar $ & $~~~~~t_c~~$ & $~~~t_d$ &$~~~D_{quan}/D_{clas} $ \\
  \hline
  $8\times 10^{-5}$ & ~~~$95\pm 30$~  & $~~~>10^4$& $~~~0.85\pm 0.05$ \\
  $1\times 10^{-4}$ & ~~~$64\pm 25$~  & $~~~>10^4$& $~~~0.74\pm 0.04$ \\
  $2\times 10^{-4}$ & ~~$22\pm 6$~  & $~~~>10^4$& $~~~0.43\pm 0.04$  \\
  $5\times 10^{-4}$ & ~~$7\pm 3$ & $~~~>10^4$ & $~~~0.07\pm 0.02$  \\
  $1\times 10^{-3}$ & ~~$3\pm 1$ & $~~~>10^4$ & $~~~0.02\pm 0.01$ \\
  $2\times 10^{-3}$ & ~~$3\pm 1$ & $ ~~~80\pm 30$ & $~~~0.02\pm 0.01$  \\
  $3\times 10^{-3}$ & ~~$3\pm 1$ & $ ~~~50\pm 20$ & $~~~0.02\pm 0.01$ \\
  $6\times 10^{-3}$ & ~~$3\pm 1$ & $ ~~~4\pm 2$ & $~~~-$ \\
  \hline
\end{tabular}
\caption{Breaking time $t_c$, dynamical localization time $t_d$,
and classical/quantum diffusion coefficient against $\hbar$ for
$K=0.2,\epsilon=0.25$, and $\tau=10^4$. $D_{clas}\approx
1.83\times 10^{-5}$, which is evaluated over $t<t_{leak}\approx
200$. $D_{quan}$ is evaluated over the time range $t_c<t<t_{leak}$
for $\hbar\le \hbar_c\approx 10^{-3}$ and $t_c<t<t_d$ otherwise.}
\label{aa}
\end{table}

For longer times $t > t_d$ diffusion stops as a consequence of
standard dynamical localization similar to the one observed in the
single quantum kicked rotor. Our results thus suggest that in
order to observe genuine quantum anomalous diffusion the value of
$\hbar$ must be such that $t_d \gg t_c$. This corresponds with
$\hbar \leq \hbar_{c}$. We remark that this condition should be
met by any experiment aiming to confirm the results reported in
this paper.

Although not shown, we have confirmed that the dependence of the
classical diffusion constant is $D \sim K^3$ similar to the case
of a single kick rotor in a region populated by classical cantori.
This together with the anomalous dependence with $\hbar$ also
found in studies of the role of cantori in quantum mechanics
\cite{Maitra} is a further indication that the $2\delta-$KR in the
region of parameters studied in this paper can be utilized as a
effective model to investigate the role of classical cantori in
quantum mechanics.

Finally we note that $\langle p^2(t) \rangle \sim t$ is only a
necessary condition for normal diffusion but this by no means
assures that the density of probability is Gaussian-like
\cite{klafter}.  Indeed, in the next section we will show that the
quantum and classical density of probability of our model strongly
deviate from the normal diffusion prediction.

\subsection{Density of probability}
The density of probability of finding a particle with momentum $p$
after a time $t$ from a given initial state
$|\psi(0)\rangle=|l_0\rangle$ is given by $P_q(p,t)\equiv
P_q(k,t)=|\langle k|\psi(t)\rangle |^2$ with $p=k\hbar$. The
parameter set $K=0.2$, $\epsilon = 0.05$ and $\tau = 10^4$ chosen
permits us to study generic features of the motion. By generic we
mean the trapping region is large enough to be studied and the
classical phase space has not stability island; namely, $K\epsilon
\ll 1$. In addition $\tau \gg 1$, so consecutive pairs of kicks
are uncorrelated. In all cases our initial state is located in the
trapping region.

\subsubsection{Classical density of probability}

The classical $P(p,t)$ is obtained by evolving the classical
equation of motion for $2\times 10^7$ different initial conditions
at the center of the trapping region. Positions are uniformly
distributed along the interval $(-\pi,\pi)$.

$P(p,t)$ is approximated as the set of the states of the whole
ensemble that falls in $(p-\Delta p/2,p+\Delta p/2)$ at time $t$.
In the numerical calculations we set $\Delta p=0.02 \sqrt{\langle
p^2(t)\rangle}$.

We distinguish the following regions in the classical density of
probability (see Fig. 2a): For short times such that the particle
is well inside the trapping region the diffusion is normal and
$P(p,t)$ is Gaussian (see inset of Fig. 2a). As the boundary of
the trapping region is approached we observe a gradual crossover
from normal to anomalous (super) diffusion. For small momentum
$P(p,t)$ is still Gaussian as leaking to the outside region is
weak. As time approaches $t_{leak} \approx 200$, the typical time
to reach the edge of the trapping region, the central (small
momentum) Gaussian region becomes smaller and smaller. Meanwhile,
the outskirts bend down and a power-like behavior typical of
anomalous diffusion is observed.

\begin{figure}
\includegraphics[width=.95\columnwidth,clip]{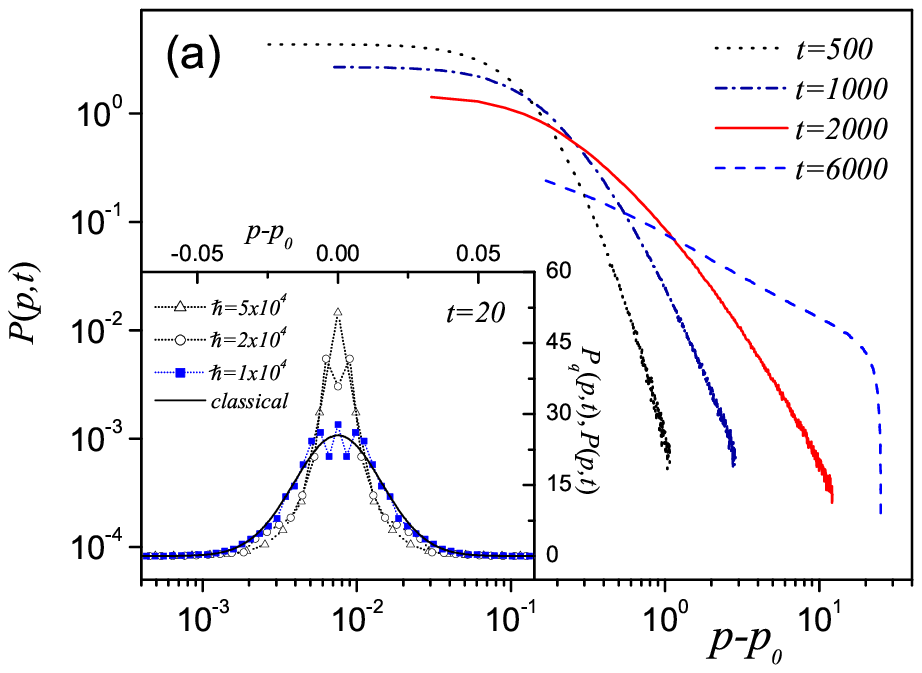}
\vspace{-.5cm}
\includegraphics[width=.95\columnwidth,clip]{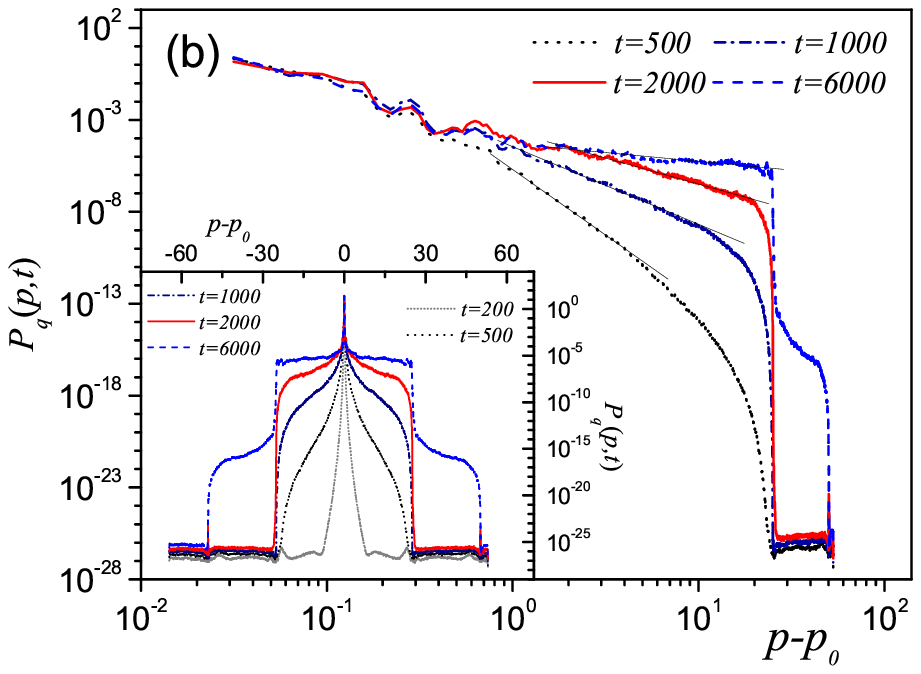}
\includegraphics[width=.95\columnwidth,clip]{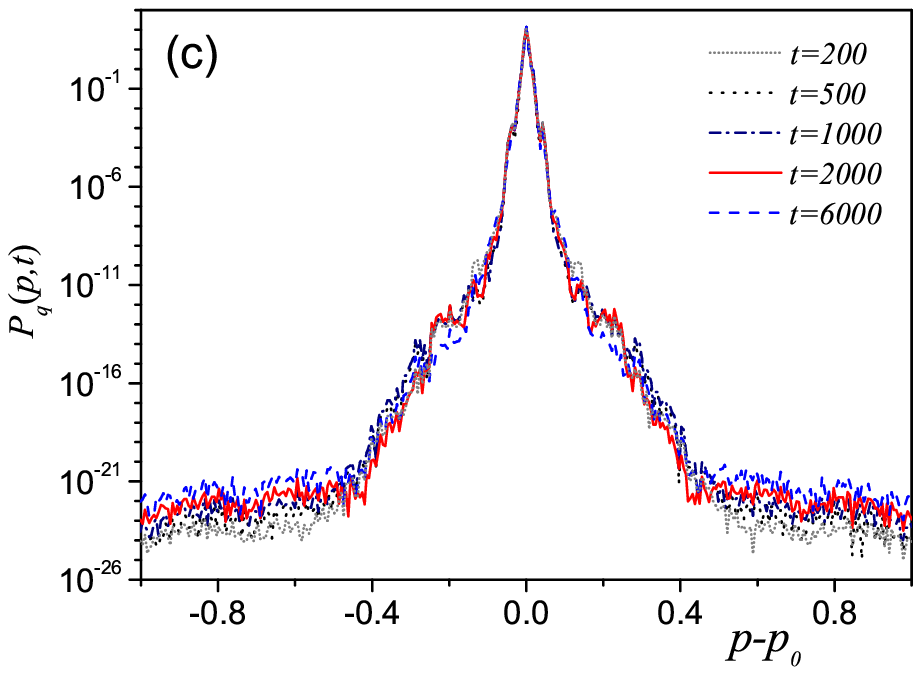}
\vspace{-.5cm} \caption{(Color online) Distribution of the
classical (a) and quantum probability density with $\hbar\approx
5\times 10^{-4}$ (b) and  $3\times 10^{-3}$ (c) respectively.
Classical $P(p,t)$ is characterized by a Gaussian profile (inset
of Fig. (a)) before leaking from the trap occurs
$(t<t_{leak}\approx 200)$. For later times $t_{leak} < t < t_d$,
$P(p,t)$ develops power-law tails, $P(p,t)\sim p^{-\alpha}$,
typical of anomalous diffusion. The exponent $\alpha$ varies in
time ($\alpha\approx 4.5,2.9,2.3,1.3$ for $t=500,1000,2000$ and
$6000$ respectively). In the quantum case for small $\hbar$
leaking leads to power-law tails as well (see Fig. (b)). The best
fitting exponent, indicated by thin black lines, is $\alpha\approx
7.1,4.2,3.0,0.94$ for $t=500,1000,2000$ and $6000$ respectively).
For larger $\hbar$ (see Fig. (c)), dynamical localization
suppresses the diffusion before leaking occurs.
$K=0.2,\epsilon=0.25$ and $\tau=10^4$.}
\end{figure}

For longer times, $t > t_{leak}$,  $P(p,t)\sim p^{-\alpha}$ with
$\alpha$ a decreasing function of time (Fig. 2a). For $t \to
\infty$, $\alpha \to 0$. This behavior is not surprising due to
the transient nature of the trapping-leaking mechanism. The
power-law decay is a direct consequence of the enhanced diffusion once
the particle has escaped from the trapped region. However,
precisely due to this fast diffusion, the particle soon approaches
a new trapped region where diffusion is slowed down again until
the particle leaks from the trap. As a consequence of this bottle
neck, the exponent $\alpha$ decreases and the density of
probability in the region of fast diffusion gradually increases
with time. In addition a sharp jump in $P(p,t)$ is observed
between the enhanced diffusion zone and the new trapping region
(see Fig. 2a for $t=6000$ where a steep drop of $P(p,t)$ happens
at around $|p-p_0|=2\pi/\epsilon$ corresponding to the new
trapping region). For long times $t \to \infty$, $P(p,t)$ resembles
a staircase, flat between trapping regions and discontinuous in
the trapping areas. This is a simple consequence of the fact that
the dwelling time in the trapping region is typically much longer
than the one needed to travel between two consecutive traps. We
shall see in the next section that quantum mechanically the
situation is different.

\subsubsection{Quantum density of probability}

The quantum density of probability $P_q(p,t)$ was calculated from
an initial condition $|\psi(0)\rangle=|l_0\rangle$ with
$l_0=\pi/\hbar_\epsilon$. For given values of $\hbar$ and
$\epsilon$ the right hand side may not be an integer. Then the
nearest value of $\hbar$ was chosen to ensure this condition. Thus
$p_0=\pi/\epsilon$ exactly for both classical and quantum
calculations. As in the classical case, $P_q(p,t)$ was evaluated
by summing up the probability of falling in a bin of width $\Delta
p$. We set $\Delta p = 0.0315$ for $\hbar\approx 5\times 10^{-4}$
(Fig. 2b) and $\Delta p = 0.006$ for $\hbar\approx 3\times
10^{-3}$ (Fig. 2c). Hence it is in fact a coarse-grained result
where part of quantum fluctuations have been suppressed. It is
more instructive to start our analysis of the quantum probability
density with a general account of the expected effect of the
classical motion on the quantum dynamics. As in the single kicked
rotor we expect to observe dynamical localization for sufficiently
long times $t > t_d$ \cite{fishman}.  However the trapping (and
eventual release) mechanism changes qualitatively the route to
dynamical localization in our model. In fact we distinguish two
different scales for full dynamical localization. For $\hbar>
\hbar_c$, full dynamical localization will occur well inside the
trapping region where the classical diffusion is normal. As a
consequence the particle will not have time to escape, ($t_d \leq
t_{leak}\approx 200$ with our parameters) and the density of
probability will be time independent and will decay exponentially
as a function of the momentum $p$.  These are typical signatures
of exponential localization similar to the one observed in a
single kicked rotor. The situation is different if $\hbar<
\hbar_{c}$ is small enough such that leaking to the enhanced
diffusion region is possible. In this case we may still observe
typical features of dynamical localization within the trapping
region. However for larger momenta the density of probability
develops time-dependent power law tails typical of anomalous
diffusion. Full dynamical localization will not typically take
place until the next trapping region is reached or later if
$\hbar$ is small enough.  This revival of quantum diffusion is a
typical feature of our model that it is not observed in the single
kicked rotor. Below we provide a more detailed account of the
relevant time and momentum scales (see Fig. 2b) for an accurate
description of the quantum density of probability:

1. For $t,p$ well within the trapping region: $t < t_{leak}$ and
$|p-p_0| < p_{leak} \sim 0.1$. After a narrow region in which
classical and quantum results fully agree (see inset Fig. 2a), we
observe in the quantum case a cusp for $p \sim p_0$ caused by the
incipient dynamical localization in the core of the trapping
region. However for larger momentum the distribution is Gaussian
in agreement with previous results (see Fig. 1) for the energy
diffusion.

2. For $p$ within the trapping region ($|p -p_0| < p_{leak}$) but
$t > t_{leak}$. The core of the quantum probability is still
exponentially localized (see Fig. 2c) but the outskirts $P_q(p,t)
\sim 1/p^\alpha$ develops a power-law tail  (see Fig. 2b) typical
of anomalous diffusion. In this region the exponent $\alpha$
depends on $\hbar$ but it is time independent.

3. For $p$ outside or close to the edge of the trapping region and
$t \geq t_{leak}$. The quantum probability $P_q(p,t) \sim
1/p^\alpha$ develops power-law tails but in this case the exponent
$\alpha$ decreases with time and eventually tends to zero for $t
\to \infty$ (see Fig. 2b). This is in principle surprising.  It
is in general expected that destructive interference slows down
the quantum motion. As a consequence the quantum density of probability should
decay faster than the classical one. However there is a simple
explanation for this behavior: in a first stage, the quantum
exponent $\alpha$ is larger than the classical one as a
consequence of destructive interference effects. For later times,
the classical probability between two trapping region become
smaller than the quantum one since the classical particle leaks
faster into the next cell. At the same time destructive
interference slows down the quantum motion in this region.
Eventually $P_q(p,t)$ saturates,  $\alpha \to 0$ before the next
trapping region is reached (see $t=6000$ in Fig. 2a and Fig. 2b).

4. For times sufficiently long $t > t_{d}$ quantum destructive
interference effects dominate, standard dynamical localization
takes place, and diffusion stops. $P_q(p,t)$ becomes time
independent and decays exponentially as a function of the
momentum. As was mentioned previously the value of $t_{d}$ depends
dramatically on whether dynamical localization occurs in the
trapping region or after it has escaped from the trapping region.
In the former case $t_d <t_{leak}\approx 200$ and in the latter
$t_d>10^4$ with no intermediate values.

\subsection{Return Probability}
Having studied diffusion in momentum for a fixed time $t$ we now
look at the explicit dependence in time. In order to proceed we
calculate return probabilities $P(t)$ of a wave-packet as a
function of time. We average over $N$ initial starting conditions
close to the center of the trapping region to suppress quantum
fluctuations,
\begin{equation}
P(t)=  \left( 1/N \right) \sum_{l_1}^{l_2} \langle P_l(t) \rangle
, \label{eq14a}
\end{equation}
where $P_l(t)= |\langle \psi(t)|\psi_l(t=0) \rangle |^2$. The
initial condition was taken to be an angular momentum eigenstate
$|\psi_l(t=0)\rangle = |l \rangle$. The results were further
averaged from $l_1\approx \pi/\hbar_\epsilon-N/2$ to $l_2=l_1+N$.
In order to ensure the initial conditions are within the trapping
region, the value $N$ is set to be $N\approx \delta p/\hbar$ with
$\delta p$ the width of the trapping region.

The return probability provides valuable information about the
degree of localization of a system. Indeed it was already utilized
in the landmark paper by Anderson \cite{anderson} about
localization. A non zero $P(t)$ in the $t \to \infty$ limit is a
signature of exponential localization. On the other hand $P(t)
\propto t^{-d/2}$ with $d$ the spatial dimensionality is typical
of fully delocalized eigenstates (normal diffusion). In the theory
of localization a power-law decay $P(t) \propto 1/t^{\gamma}$ with
$\gamma < d/2$ is a signature of localization typical of a
disordered conductor close to a localization transition. Such slow
decay has also been related to the effect of cantori in  transport
\cite{Ketz2} properties.

\begin{figure}
\includegraphics[width=.95\columnwidth,clip]{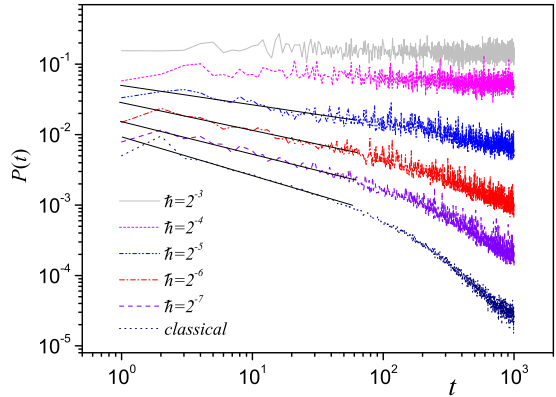}
\vspace{-.5cm} \caption{(Color online) Return probability as a
function of time for $K=5, \epsilon=0.04$ and $\tau=10^4$.  We present results 
for different values of  $\hbar$ together with the classical prediction. In the
trapping region ($(t<t_{leak}\approx 30)$ in this case) the return probability
decays as a power-law $P(t)\sim t^{-\gamma}$. The exponent
$\gamma$, given by the best linear fitting (indicated by a thin
black line), decreases as  $\hbar$ increases. We have obtained
$\gamma \approx 0.50,0.46,0.40$ and $0.28$ for the classical and
quantum cases with $\hbar=2^{-7}, 2^{-6}$ and $2^{-5}$
respectively. } \label{figure3}
\end{figure}

Our results are summarized in Fig. 3. For the sake of comparison,
we first study the classical counterpart of $P(t)$. Again $2\times
10^7$ initial states uniformly located in space and with
$p_0=\pi/\epsilon$ were evolved. $P(t)$ was evaluated as the set
of states that fall in the momentum region $|p-p_0|<0.004$ at time
$t$. For the parameters ($K=5, \epsilon=0.04$ and $\tau=10^4$)
chosen for the calculation, the width of the trapping region
$\delta p\approx 4$, and $t_{leak}\approx 30$. For $t<t_{leak}$,
$P(t)\sim t^{-0.5}$,  in agreement with the Gaussian probability
density distribution obtained in the previous section. For
$t>t_{leak}$, leaking gradually takes over and after a crossover
$P(t)$ undergoes a faster power-law decay, $P(t) \sim t^{-1.5}$
until the next trap is reached.

Within the trapping region the quantum $P(t)$ decays as
$t^{-\gamma}$ with $\gamma < 0.5$, a decreasing function of
$\hbar$.  The fact that $\gamma < d/2$ is a clear indication that
destructive interference is already slowing down the quantum
diffusion. Not surprisingly, $\gamma$ approaches its classical
limit, $0.5$, as $\hbar \to 0$.

For sufficiently large $\hbar \geq 0.1$ the particle cannot
escapes from the trapping region and $P(t)$ becomes constant (see
for example $P(t)$ for  $\hbar=2^{-3}$ in Fig. 3). This is fully
consistent with previous results for the energy diffusion and the
density of probability.

For smaller $\hbar$ the particle escapes from the trapping region
and full dynamical localization occurs only for times much longer
than those shown in Fig. 3. For $t > t_{leak}$ we observe a rapid
decrease of $P(t)$. The decay is still power-law but the exponent
is larger than for $t < t_{leak}$ though smaller than the
classical result ($\gamma\approx 1.5$). This time scale can be
utilized to study the interplay between quantum effects and
classical anomalous diffusion. Destructive interference slows down
the quantum motion however diffusion is not arrested due to the
underlying classical anomalous (super)diffusion.

Finally we address the similarities of our model with those of a
generic Hamiltonian in a region of the phase space dominated by
cantori. Qualitatively the effect of the trapping region is
similar to that of cantori \cite{Geisel,weiss}. In both cases the
particle remains in a small region of the phase space for a long
time until eventually is released to the chaotic sea. In addition
the dependence of the classical diffusion constant $D \sim K^3$
\cite{Tania} is identical in both cases. In the quantum realm
cantori \cite{borgonovi,casati,sirko} induces slow power-law decay
in quantities such as the return probability \cite{Ketz2} or the
eigenstates themselves \cite{borgonovi}. In addition cantori have
been linked to fractional scaling with $\hbar$  \cite{Maitra} and
anomalous dynamical localization \cite{zaslavs}. All these
features have also been observed in the model studied in this
paper. These similarities strongly suggest that the $2-\delta$ KR
can be utilized as a toy model to study the effect of cantori in
classical and quantum mechanics.

\section{Conclusions}

We have studied the effect of classical anomalous diffusion on the
quantum dynamics of atoms exposed to pairs of $\delta$-kicks. Our
result  are generic, do not depend on the parameters
$K\epsilon \ll 1$ and $\tau \gg 1$ utilized. We have identified a
regime of quantum diffusion where the motion is slow down due to
destructive interference but full dynamical localization has not
occurred yet. This has been related to the effect of the
underlying classical trapping and releasing mechanism on the
quantum dynamics. We have argued that the $2\delta$-KR can be used
as a simplified model to study the effect of cantori in classical
and quantum mechanics.

\acknowledgments

The authors thank C.E. Creffield and T.S. Monteiro for helpful
discussions. AMG acknowledges financial support from a Marie Curie
Outgoing Action, contract MOIF-CT-2005-007300. JW acknowledges
support from Defence Science and Technology Agency (DSTA) of
Singapore under agreement of POD0613356.

\end{document}